\documentclass[twocolumn,,amsmath,amssymb]{revtex4}

\usepackage{graphicx}
\usepackage{dcolumn}
\usepackage{bm}
\newcommand{\etal}{\emph{et al.}}
\newcommand{\be}{\begin{equation}}
\newcommand{\ee}{\end{equation}}
\newcommand{\bfig}{\begin{figure}}
\newcommand{\efig}{\end{figure}}

\begin{document}      

\title{Experimental tests of the chiral anomaly magnetoresistance in the Dirac-Weyl semimetals Na$_3$Bi and GdPtBi
} 
\noindent

\author{Sihang Liang$^1$}
\author{Jingjing Lin$^1$}
\author{Satya Kushwaha$^2$}
\author{Jie Xing$^3$}
\author{Ni Ni$^3$}
\author{R. J. Cava$^2$}
\author{N. P. Ong$^{1,\dagger}$}
\affiliation{
$^1$Department of Physics, $^2$Department of Chemistry, Princeton University, Princeton, NJ 08544\\
$^3$Department of Physics and Astronomy and California NanoSystems Institute, University of California, Los Angeles, CA 90095
}

\date{\today}      
\begin{abstract} 
In the Dirac/Weyl semimetal, the chiral anomaly appears as an ``axial'' current arising from charge-pumping between the lowest (chiral) Landau levels of the Weyl nodes, when an electric field is applied parallel to a magnetic field $\bf B$. Evidence for the chiral anomaly was obtained from the longitudinal magnetoresistance (LMR) in Na$_3$Bi and GdPtBi. However, current jetting effects (focussing of the current density $\bf J$) have raised general concerns about LMR experiments. Here we implement a litmus test that allows the intrinsic LMR in Na$_3$Bi and GdPtBi to be sharply distinguished from pure current jetting effects (in pure Bi). Current jetting enhances $J$ along the mid-ridge (spine) of the sample while decreasing it at the edge. We measure the distortion by comparing the local voltage drop at the spine (expressed as the resistance $R_{spine}$) with that at the edge ($R_{edge}$). In Bi, $R_{spine}$ sharply increases with $B$ but $R_{edge}$ decreases (jetting effects are dominant). However, in Na$_3$Bi and GdPtBi, both $R_{spine}$ and $R_{edge}$ decrease (jetting effects are subdominant). A numerical simulation allows the jetting distortions to be removed entirely. We find that the intrinsic longitudinal resistivity $\rho_{xx}(B)$ in Na$_3$Bi decreases by a factor of 10.9 between $B$ = 0 and 10 T. A second litmus test is obtained from the parametric plot of the planar angular magnetoresistance. These results strenghthen considerably the evidence for the intrinsic nature of the chiral-anomaly induced LMR. We briefly discuss how the squeeze test may be extended to test ZrTe$_5$.
\end{abstract}

\maketitle      
\section{Introduction}\label{intro}
In the past 2 decades, research on the Dirac states in graphene and topological insulators has uncovered many novel properties arising from their linear Dirac dispersion. In these materials, the Dirac states are confined to the two-dimensional (2D) plane. Interest in three-dimensional (3D) Dirac states may be traced to the even earlier prediction of Nielsen and Ninomiya (1983)~\cite{Nielsen} that the chiral anomaly may be observable in crystals (the space-time dimension 3+1D needs to be even). The anomaly, which appears as a current in a longitudinal magnetic field $\bf B$, arises from the breaking of a fundamental, classical symmetry of massless fermions, the chiral symmetry. Recent progress in topological quantum matter has led to several systems that feature protected 3D Dirac and Weyl states in the bulk~\cite{Kane,Ashvin}.

\begin{figure*}[t]
\includegraphics[width=14 cm]{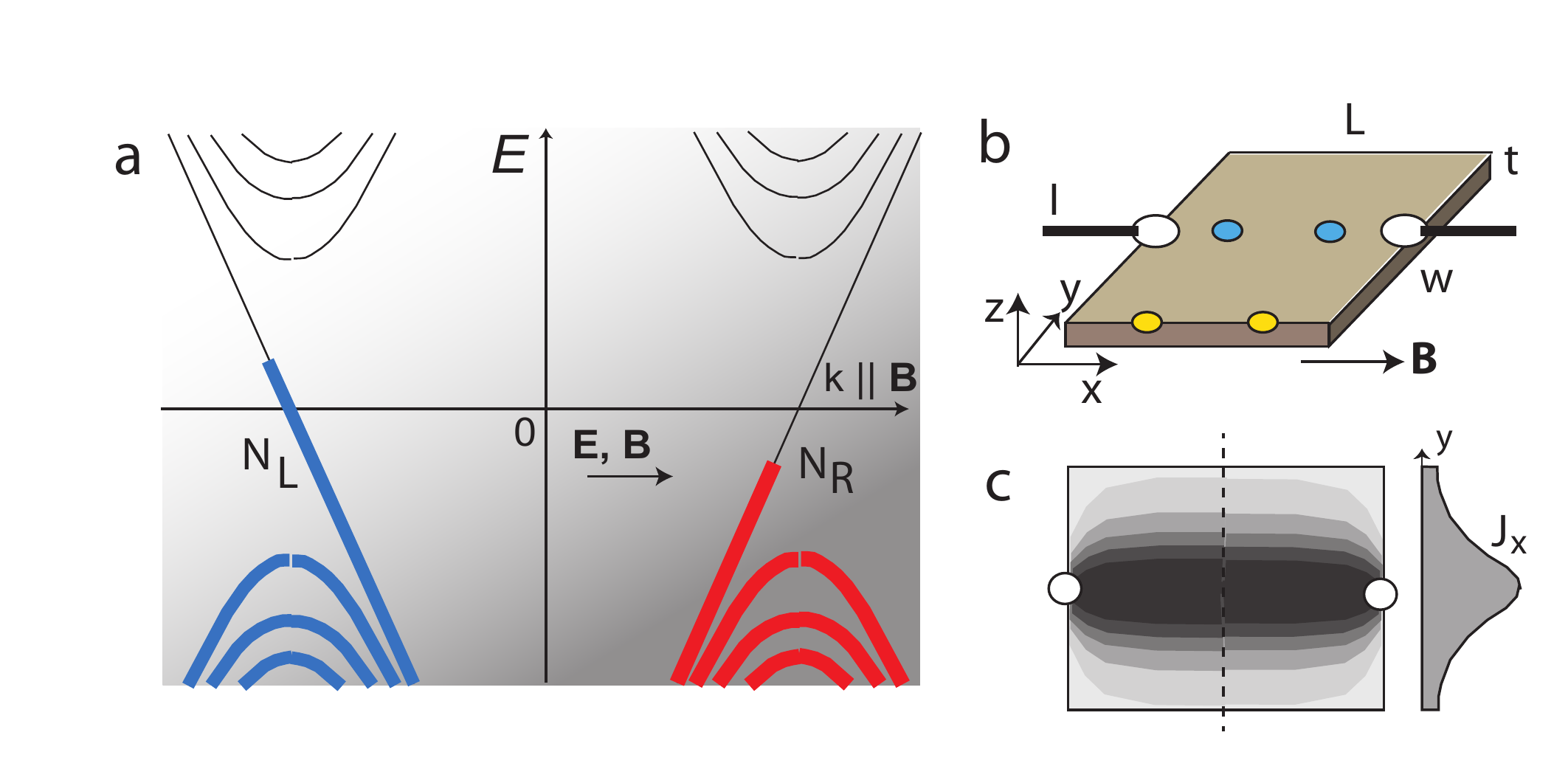}
\caption{\label{figLandau} 
The Landau spectrum of Weyl fermions (Panel (a)). In a field $\bf B\parallel \hat{x}$, the lowest Landau levels are chiral with velocity $\bf v$ either $\parallel {\bf B}$ or -$\bf B$. Application of $\bf E\parallel B$ transfers charge between them, which increases the left-moving population $N_L$ at the expense of the right-moving population $N_R$ (the bold blue and red curves indicate occupations of the LLs). Panel (b) shows a pair of voltage contacts (blue dots) placed on the line joining the current contacts (white circles). A second pair (yellow) is placed along an edge. Panel (c) is a schematic drawing of the intensity map of $J_x$ (with dark regions the most intense) when current jetting effects are pronounced. The profile of $J_x$ vs. $y$ (with $x$ at the dashed line) is sketched on the right.
}
\end{figure*}

A crucial step in the search for 3D Dirac states was the realization that inclusion of point group symmetry (with time-reversal (TR) symmetry and inversion symmetry) allows Dirac nodes to be protected anywhere along symmetry axes, instead of being pinned to TR-invariant momenta on the Brillouin zone surface~\cite{Bernevig,Nagaosa}. Relaxation of this constraint led to the discovery of Na$_3$Bi~\cite{Wang1}  and Cd$_3$As$_2$, in which the 2 Dirac nodes are protected by $C_3$ and $C_4$ symmetry, respectively. In the absence of $\bf B$, each Dirac node is described by a $4\times 4$ Hamiltonian that can be block-diagonalized into two $2\times 2$ Weyl blocks with opposite chiralities ($\chi = \pm 1$). The absence of mixing between the two Weyl fermions expresses the existence of chiral symmetry. In a strong $\bf B$ the Weyl states are quantized into Landau levels. As shown in Fig. \ref{figLandau}a, a distinguishing feature is that the lowest Landau level (LLL) in each Weyl node is chiral with a velocity $\bf v$ strictly $\parallel\bf B$ (or $-\bf B$ as dictated by $\chi$)~\cite{Ashvin}.

As a result, electrons occupying the LLL segregate into two massless groups -- left- and right-movers with populations $N_L$ and $N_R$, respectively. Independent conservation of $N_L$ and $N_R$ implies that the chiral charge density $\rho^5 = (N_L-N_R)/V$ is conserved, just like the total charge density $\rho_{tot} = (N_L+N_R)/V$ ($V$ is the sample volume). 
However, application of an electric field $\bf E\parallel B$ breaks the chiral symmetry by inducing mixing between the left- and right-moving branches (Fig. \ref{figLandau}a) (for a pedagogical discussion, see Ref.~\cite{Rubakov}). A consequence is that conservation of $\rho^5$ is violated by a quantity ${\cal A}$ called the anomaly term, viz. $\nabla \cdot {\bf J}^5 + \partial_t \rho^5 = e{\cal A}$, where ${\bf J}^5$ is the axial current density. (From the density of states in the LLL and the rate of change $\partial k_z/\partial t$ induced by $\bf E$, we obtain ${\cal A} = (e^2/4\pi^2\hbar^2){\bf E\cdot B}$.) The presence of ${\bf J}^5$ is detected as a large, negative, longitudinal magnetoresistance (LMR). This constitutes the chiral anomaly. [The anomaly first appeared in the theory of $\pi$-meson decay~\cite{Adler,Bell}. See Refs. \cite{Peskin,Bertlmann}.] The conditions for observing the anomaly in Dirac semimetals were discussed, for e.g., in Refs. \cite{Ashvin,Ran,Spivak,Parameswaran,Burkov15}.

In 2015, Xiong \etal~reported the observation of a 5 to 6-fold suppression of the LMR in Na$_3$Bi, identifed with the chiral anomaly~\cite{Xiong}. A year later, Hirschberger \etal~\cite{Hirschberger} reported observing the chiral anomaly including its thermoelectric signature in the half-Heusler GdPtBi. Although the low-lying states in GdPtBi are not Dirac-like in zero $B$, the application of a Zeeman field splits both conduction and valence bands to produce protected crossings which define Weyl nodes. With $\bf B\parallel E$, a 5-fold LMR was observed with a profile very similar to that in Na$_3$Bi. In both Na$_3$Bi and GdPtBi, the carrier mobility is relatively low (3,000 and 2,000 cm$^2$/Vs at 2 K, respectively).

There have also been several reports of negative LMR observed in the Weyl semimetals TaAs, NbP and analogs~\cite{XiDai,Shuang,Hassinger,ZhuanXu}. However, the weakness of their LMR signals (50-100$\times$ weaker than in Na$_3$Bi) and their fragility with respect to placement of contacts, together with the high mobilities of the Weyl semimetals (150,000 to 200,000 cm$^2$/Vs) have raised concerns about current jetting artifacts~\cite{Hassinger,ZhuanXu}. As a consequence, there is considerable confusion and uncertainties about LMR experiments in general, and the LMR reported in the Weyl semimetals in particular. The concerns seem to have spread to Na$_3$Bi and GdPtBi as well, notwithstanding their much larger LMR signal.

There is good reason for the uncertainties. Among the resistivity matrix elements, measurements of the longitudinal resistivity $\rho_{xx}$ (for $\bf B\parallel\hat{x}$) are the most vulnerable to inhomogeous flow caused by current jetting. Even when the LMR signal in a sample is mostly intrinsic, the chiral anomaly produces an intrinsic conductivity anisotropy $u$ which unavoidably produces inhomogeneous current distributions that distort the observed LMR profile. Given the prominent role of LMR in chiral-anomaly investigations, it is highly desirable to understand these effects at a quantitative level, and to develop a procedure that removes the distortions.

A major difference between the large LMR systems Na$_3$Bi and GdPtBi on the one hand and the Weyl semimetals on the other is their carrier densities. Because the density is low in both Na$_3$Bi ($1\times 10^{17}$ cm$^{-3}$) and in GdPtBi ($1.5\times 10^{17}$ cm$^{-3}$), the field $B_Q$ required to force the chemical potential $\zeta$ into the LLL is only 5-6 T. By contrast, $B_Q$ is 7-40 T in the Weyl semimetals. As shown in Fig. \ref{figLandau}a, the physics underlying the anomaly involves the occupation of chiral, massless states. Occupation of the higher LLs (when $B<B_Q$) leads to strong suppression of the anomaly~\cite{Hirschberger}. Moreover, as discussed below, LMR measurements involve a competition between the anomaly mechanism (``the quantum effect'') and classical current jetting effects which onset at a second field scale $B_{cyc}$. The relative magnitudes of these field scales dictate which effect dominates.

Here we report a series of experiments designed to separate intrinsic from extrinsic effects in LMR experiments. Focussing of the current density $J({\bf r})$ into a beam strongly reduces its value at the edges of a sample. As shown in Sec. \ref{squeeze}, the effects of current-jetting can be neatly factored into a quantity ${\cal L}_y$ (line-integral of $J_x$) which can be measured by local voltage contact pairs. By comparing local voltage drops at the maximum and minimum of the profile of $J$, we devise a litmus test that sharply distinguishes the two chiral-anomaly semimetals from the case of pure Bi (Sec. \ref{squeeze results}). Adopting a quantitative treatment (Sec. \ref{intrinsic}) we show how the intrinsic $\rho_{xx}(B)$ can be derived from the local voltage results. Applying this technique to Na$_3$Bi, we obtain the intrinsic profiles of $\rho_{xx}(B)$ and the anisotropy, with current jetting distortions removed. The degree of distortion at each $B$ value becomes plainly evident. The competition between the quantum and classical effects is described in Sec. \ref{testscope}. To look beyond Na$_3$Bi and GdPtBi, we discuss how the tests can be extended using focussed ion beam techniques to test ZrTe$_5$, which grows as a narrow ribbon. The Weyl semimetals e.g. TaAs require availability of ultrathin films. The planar angular magnetoresistance (Sec. \ref{AMR}) provides a second litmus test -- one that is visually direct when displayed as a parametric plot (Sec. \ref{param}). In Sec. \ref{conclude}, we summarize our results.

\section{The squeeze test}\label{squeeze}
Current jetting refers to the focussing of the current density $\bf J(r)$ into a narrow beam $\parallel \bf B$ arising from the field-induced anisotropy $u$ of the conductivity (the drift of carriers transverse to $\bf B$ is suppressed relative to the longitudinal drift). To maximize the gradient of $J$ along the $y$-axis, we select plate-like samples with $L$,$w\gg t$, where $w$, $L$ and $t$ are the width, length and thickness, respectively (Fig. \ref{figLandau}b). The $x$- and $y$-axes are aligned with the edges and ${\bf B}\parallel\bf\hat{x}$. As sketched in Fig. \ref{figLandau}c, the profile of $J_x$ vs. $y$ is strongly peaked at the center of the sample and suppressed towards the edges. In the squeeze test, we measure the voltage difference across a pair of contacts (blue dots) along the line joining the current contacts (which we call the spine) as well as that across a pair on the edge (yellow dots). To accentuate current jetting effects, we keep the current contact diameters $d_c$ small ($d_c\ll w$) and place them on the top face of the sample wherever possible. (The squeeze test cannot be applied to needle-like crystals). 

\emph{Sample Preparation}
We provide details on the preparation of Na$_3$Bi which is by far the most difficult of the 3 materials to work with. Na$_3$Bi crystallizes to form hexagonal platelets with the broad face normal to (001). The crystals investigated here were grown under the same conditions as the samples used in Xiong \etal~\cite{Xiong}; they have carrier densities $1\times 10^{17}$ cm$^{-3}$ and $B_Q$ in the range 5-6 T. These crystals should be distinguished from an earlier batch~\cite{XiongEPL} that have much higher carrier densities (3-6$\times 10^{19}$ cm$^{-3}$) for which we estimate $B_Q\sim$100 T. No evidence for negative LMR was obtained in the highly-doped crystals~\cite{XiongEPL}.

Because of the high Na content, crystals exposed to ambient air undergo complete oxidation in $\sim$5 s. The stainless growth tubes containing the crystals were opened in an argon glove box equipped with a stereoscopic microscope, and all sample preparation and mounting were performed within the glove box. The crystals have the ductility of a soft metal. Using a sharp razor, we cleaved the bulk crystal into platelets $1\times 1$ mm$^2$ on a side with thickness 100 $\mu$m. Current and voltage contacts were painted on using silver paint (Dupont 4922N). A major difficulty was achieving low-resistance contacts on the top face (for measuring $R_{spine}$). After much experimentation, we found it expedient to remove a thin layer of oxide by lightly sanding with fine emery paper (within the glove box). The sample was then placed inside a capsule made of G10 epoxy. After sealing the lid with stycast epoxy, the capsule was transferred to the cryostat.

We contrast two cases. In Case (1), the anisotropy $u = \sigma_{xx}/\sigma_{yy}$ increases in a longitudinal field $\bf B$ because the transverse conductivity $\sigma_{yy}$ decreases steeply (as a result of cyclotronic motion) while $\sigma_{xx}$ is unchanged in $B$. With $\bf B\parallel\hat{x}$, the two-band model gives the resistivity matrix 
\be
\tilde{\rho}(B) = \left[
		\begin{array}{cc}
		[\sigma_e + \sigma_h]^{-1}	&	0 \\
			0								& [\frac{\sigma_e}{\Delta_e} + \frac{\sigma_h}{\Delta_h}]^{-1}
		\end{array}
		\right]
\label{rhoab}
\ee
(we suppress the $z$ component for simplicity).
The zero-$B$ conductivities of the electron and hole pockets are given by $\sigma_e = ne\mu_e$ and $\sigma_h = pe\mu_h$, with $n$ and $p$ the electron and hole densities, respectively, and $e$ the elemental charge. $\mu_e$ and $\mu_h$ are the mobilities in the electron and hole pockets, respectively, and $\Delta_e = (1+\mu_e^2B^2)$ and $\Delta_h = (1+\mu_h^2B^2)$. With $\bf B\parallel \hat{x}$, the off-diagonal elements vanish. In Case (1), we assume that $\sigma_e$ and $\sigma_h$ remain constant. Hence the observed resistivity $\rho_{xx}$ is unchanged in $B$. However, the transverse conductivity $\sigma_{yy}$ decreases  (as $1/B^2$ in high $B$). The anisotropy arises solely from the suppression of the conduction transverse to $\bf B$ by the cyclotron motion of both species of carriers. 

Case (2) is the chiral anomaly regime in the Dirac semimetal. Charge pumping between Landau levels (LLs) of opposite chirality leads to an axial current which causes $\sigma_{xx}$ to increase with $B$. Simultaneously, the 1D nature of the LL dispersion suppresses the transverse conductivity $\sigma_{yy}$. Hence the increase in $u$ derives equally from the opposite trends in $\sigma_{xx}$ and $\sigma_{yy}$. 

Denoting field-induced changes by $\Delta$, we have
\begin{eqnarray}
&\mathrm{Case (1):}\;\;\Delta u >0 & \Longleftrightarrow \Delta\sigma_{xx}\sim 0; \Delta\sigma_{yy}<0, \label{R1}\\
&\mathrm{Case (2):}\;\;\Delta u >0 & \Longleftrightarrow \Delta\sigma_{xx}> 0; \Delta\sigma_{yy}<0.
\label{R2}
\end{eqnarray}

In the test, the voltage drops $V_{spine}$ and $V_{edge}$ are given by
\begin{eqnarray}
V_{edge}(B) &=& -\rho_{xx}(B) \int^\ell_0 J_x(x,w/2;B)dx \equiv \rho_{xx}{\cal L}_e, \label{edge}\\ 
V_{spine}(B) &=& -\rho_{xx}(B) \int^\ell_0 J_x(x,0;B)dx \equiv \rho_{xx}{\cal L}_s\label{spine}
\end{eqnarray}
where ${\cal L}_e(B)$ and ${\cal L}_s(B)$ are the line integrals of $J_x$ along the edge and spine, respectively ($\ell$ is the spacing between voltage contacts). The intrinsic $B$ dependence (expressed in $\rho_{xx}(B)$) has been cleanly separated from the extrinsic $B$ dependences of ${\cal L}_e(B)$ and ${\cal L}_a(B)$, which arise from current focussing effects.

The area under the curve of $J_x$ vs. $y$ is conserved, i.e.
\be
\int^{w/2}_{w/2} J_x(x,y;B)dy = I,
\label{Jx}
\ee
with $I$ the applied current. At $B$ = 0, we may take ${\bf J}$ to be uniform with the magnitude $J_0 = I/(wt)$. The line integral reduces to ${\cal L}_0 = J_0\ell$. In finite $B$, focussing of the current beam implies that the current density is maximum along the spine and minimum at the edge, i.e. $J_x(x,0;B)>J_x(x,w/2;B)$. Moreover, Eq. \ref{Jx} implies that $J_x(x,0;B)>J_0>J_x(x,w/2;B)$. Hence the line integrals satisfy the inequalities 
\be
{\cal L}_s(B)>{\cal L}_0 > {\cal L}_e(B).
\label{calL}
\ee

If both $\sigma_e$ and $\sigma_h$ are $B$-independent, as in Case (1), we have from Eqs. \ref{spine}, \ref{edge} and \ref{calL},
\be
V_{spine}(B)> V_0 > V_{edge}(B),
\label{calL}
\ee
where $V_0$ is the voltage drops across both pairs at $B$ = 0. Clearly, $V_{spine}$ increases monotonically with $B$ while $V_{edge}$ decreases. Physically, focussing the current density along the spine increases the local $E$-field there. Current conservation then requires $J_x$ to be proportionately suppressed along the edges. Measuring $V_{edge}$ alone yields a negative LMR that is spurious.

In Case (2), however, $\rho_{xx}$ decreases intrinsically with $B$ because of the chiral anomaly while ${\cal L}_s$ increases. Competition between the 2 trends is explicitly seen in the profile of $V_{spine}$ vs. $B$. 
As shown below, in Na$_3$Bi and GdPtBi, the intrinsic decrease in $\rho_{xx}$ dominates so both $V_{spine}$ and $V_{edge}$ decrease with $B$. We remark that, by Eq. \ref{calL}, $V_{spine}$ always lies above $V_{edge}$. Moreover, when the rate of increase in ${\cal L}_s$ begins to exceed (in absolute sense) the rate of decrease in $\rho_{xx}$ at sufficiently large $B$, the curve of $V_{spine}(B)$ can display a broad minimum above which $V_{spine}$ increases.

Hence, if both $V_{spine}$ and $V_{edge}$ are observed to decrease with increasing $B$, the squeeze test provides positive confirmation that the observed LMR is intrinsic. Their field profiles bracket the intrinsic behavior of $\rho_{xx}$. Conversely, if intrinsic LMR is absent (i.e. $\sigma_{xx}$ is unchanged), $V_{spine}$ and $V_{edge}$ display opposite trends (the marginal case when the intrinsic LMR is weak is discussed in Sec. \ref{intrinsic}). 

We remark that the current jetting effects cannot be eliminated by using very small samples (e.g. using nanolithography). As long as we remain in the classical transport regime, the equations determining the functional form of $J(x,y)$ in strong $B$ are scale invariant. Because intrinsic length scales (e.g. the magnetic length $\ell_B$ or the skin depth $\delta_s$) are absent in classical dc transport, the same flow pattern is obtained on either mm or micron-length scales.

\section{Results of squeeze test}\label{squeeze results}
\begin{figure*}[t]
\includegraphics[width=16 cm]{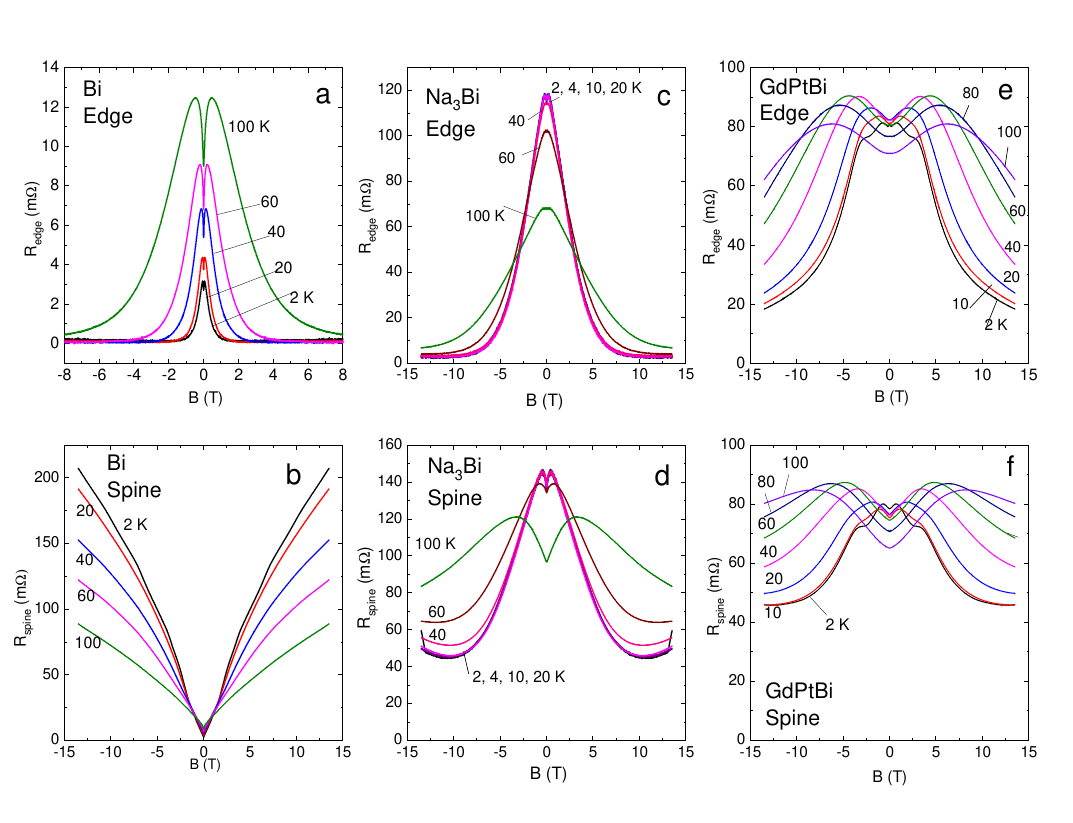}
\caption{\label{figspine} 
Comparison of squeeze-test results in pure bismuth (Case (1)) and in Na$_3$Bi and GdPtBi (both Case (2)). Panels (a) and (b) display the voltage drops $V_{edge}$ and $V_{spine}$ (expressed as $R_{edge}$ and $R_{spine}$, respectively) measured in Bi (Sample B1). In Panel (a), $R_{edge}$ displays a steep decrease with increasing $B$ that steepens as $T$ decreases from 100 K to 2 K. By contrast, $R_{spine}$ increases steeply. The opposite trends in (a) and (b) implies that both arise from pure current jetting, strictly reflecting changes in ${\cal L}_e$ and ${\cal L}_s$, respectively (Eqs. \ref{Re} and \ref{Rs}). By contrast, in Na$_3$Bi (Sample N1), both $R_{edge}$ decrease with $B$ (Panels (c) and (d), respectively). This implies that $\rho_{xx}(B)$ decreases uniformly throughout the sample. Nonetheless, current focussing effects (expressed by ${\cal L}_e$ and ${\cal L}_s$) are visible. Below 20 K, ${\cal L}_e$ exaggerates the decrease in $R_{edge}$ while ${\cal L}_s$ counters some of the intrinsic decrease in $R_{spine}$. A telling feature is the weak upturn in $R_{spine}$ above 10 T. When the intrinsic LMR is saturated, ${\cal L}_s$ produces a weak increase in $R_{spine}$. In GdPtBi (Sample G1, Panels (e) and (f)), both $R_{edge}$ and $R_{spine}$ also decrease with increasing $B$. Below 10 K, the decrease in $R_{edge}$ is much larger than that in $R_{spine}$. The latter also attains a broad minimum above 10 T (but not the former). These features confirm that ${\cal L}_e$ amplifies the growth in the intrinsic LMR while ${\cal L}_s$ partially counters it.
}
\end{figure*}

The results of applying the squeeze test on the 3 systems are summarized in Fig. \ref{figspine}. In Panels (a) and (b) we show the voltage drops $V_{edge}$ and $V_{spine}$ measured in pure bismuth (Sample B1). The signals are expressed as the effective resistances 
\begin{eqnarray}
R_{edge} &=& \rho_{xx}(B).{\cal L}_e(B)/I, \label{Re}\\
R_{spine} &=& \rho_{xx}(B).{\cal L}_s(B)/I \label{Rs}.
\end{eqnarray}
The steep decrease in $R_{edge}$ (Panel (a)) illustrates how an apparent but spurious LMR can easily appear when the mobility is very high (in Bi, $\mu_e$ exceeds $10^6$ cm$^2$/Vs at 4 K). Comparison of $R_{edge}$ and $R_{spine}$ measured simultaneously shows that they have opposite trends vs. $B$. As $\sigma_e$ and $\sigma_h$ are obviously $B$ independent in Bi, $\rho_{xx}$ is also $B$ independent. By Eqs. \ref{Re} and \ref{Rs}, the changes arise solely from ${\cal L}_e$ and ${\cal L}_s$. Hence, Figs. \ref{figspine}a and \ref{figspine}b verify experimentally that $V_{edge}$ and $V_{spine}$ display the predicted large variations of opposite signs when current jetting is the sole mechanism present (see simulations in Sec. \ref{intrinsic}).

Next we consider Na$_3$Bi. In this sample (N1), $R_{edge}$ below 20 K decreases by a factor of 50 between $B$ = 0 and 10 T (Fig. \ref{figspine}c). This is an order of magnitude larger than observed in Ref. \cite{Xiong}. The increase arises from enhanced current focussing effect in the present contact placement utilizing small current contacts attached to the broad face of the crystal, as well as a larger $u$. In spite of the enhanced jetting, $R_{spine}$ shows a pronounced decrease is contrast to the case for Bi. The intrinsic decrease in $\rho_{xx}$ dominates the increase in ${\cal L}_s$ throughout (see Eq. \ref{Rs}). Hence we conclude that there exists a large intrinsic, negative LMR that forces $R_{spine}$ to decrease despite focussing of $J({\bf r})$ along the ridge. Further evidence for the competing scenario comes from the weak minimum at 10 T in the curves below 40 K in Fig. \ref{figspine}d. As anticipated above, in large $B$, $\rho_{xx}$ approaches a constant because of the saturation chiral-anomaly term. However, ${\cal L}_s$ continues to increase because the transverse conductance worsens. Consequently, $R_{spine}$ goes through a minimum before increasing. This is seen in $R_{spine}$, but absent in $R_{edge}$.

A feature that we currently do not understand is the large $V$-shaped cusp in weak $B$. At 100 K, the cusp is prominent in $V_{spine}$ but absent in $V_{edge}$. 

In Fig. \ref{figspine}(e) and (f), we show the field profiles of $R_{edge}$ and $R_{spine}$ measured in GdPtBi (Sample G1). Again, as in Na$_3$Bi, the anomaly-induced decrease in $\rho_{xx}$ dominates the increase in ${\cal L}_s$, and $R_{spine}$ is observed to decrease in increasing $B$. 
The relative decrease in $R_{edge}$ is larger than that in $R_{spine}$. Further, $R_{spine}$ below 10 K shows the onset of a broad minimum above 10 T (Fig. \ref{figspine}f), whereas $R_{edge}$ continues to fall. 

These features, in accord with the discussion above, are amenable to a quantitative analysis that yields the intrinsic field profiles of both $\rho_{xx}$ and $u$ (next section).

\section{The intrinsic LMR profile}\label{intrinsic}
\begin{figure*}[t]
\includegraphics[width=16 cm]{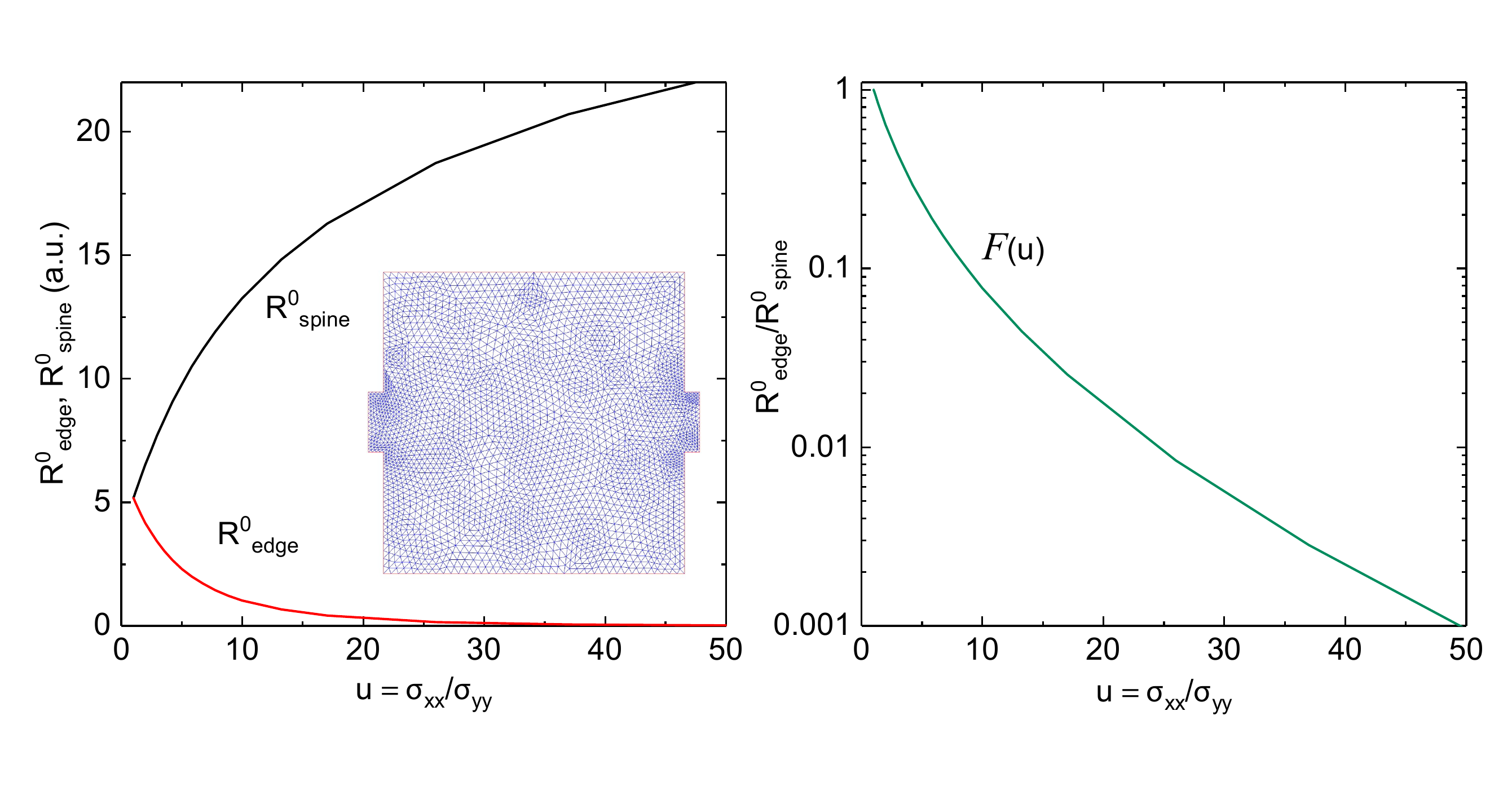}
\caption{\label{figsimulation} 
Procedure to obtain the intrinsic LMR curve $R_{intr}$ vs. $B$ and the intrinsic anisotropy $u = \rho_{yy}/\rho_{xx}$ from measurements of $R_{edge}$ and $R_{spine}$. Panel (a) plots the calculated resistances $R^0_{edge}$ and $R^0_{spine}$ vs. $u$ with $\rho_{xx}=\rho_0$ (Case (1)) (obtained by solving the anisotropic Laplace equation (Eq. \ref{eq:Laplace}) in a 2D sample (matched to $x$ and $y$ dimensions of the sample). The divergent trends reflect the enhancement of $J({\bf r})$ along the spine and its decrease at the edge. The inset shows the triangulation network generated during the simulation. The stubs on the left and right edges are current contacts. Panel (b) plots the calculated template function ${\cal F}(u)$ in semilog scale (Eq. \ref{eq:F}). Equating the measured ratio ${\cal G}(B)$ to ${\cal F}(u)$, and using Eq. \ref{eq:u}, we derive the intrinsic field profiles of $R_{intr}$ and $u$ (see Fig. \ref{figintrinsic}). 
}
\end{figure*}

The factorization expressed in Eqs. \ref{Re} and \ref{Rs} allows us to obtain the intrinsic field profile of $\rho_{xx}(B)$ in the face of strong current-density inhomogeneity induced by jetting. To start, we note that, once the boundaries are fixed, the functional form of the inhomogeneous current density ${\bf J(r)}$ depends only on the conductivity anisotropy $u$ regardless of its microscopic origin. Assuming a constant $\rho_{xx} = \rho_0$ (i.e. Case (1)), we first calculate by numerical simulation the effective resistances $R^0_{edge}(u)$ and $R^0_{spine}(u)$ over a broad range of $u$. For simplicity, the simulation is performed for a sample in the 2D limit by solving the anisotropic Laplace equation 
\be
[\sigma_{xx}\partial_x^2 + \sigma_{yy}\partial_y^2 ]\psi(x,y) = 0
\label{eq:Laplace}
\ee
for the potential $\psi(x,y)$ at selected values of $u$. We used the relaxation method on a triangular mesh with Dirichlet boundary conditions at the current contacts (inset in Fig. \ref{figsimulation}a). 

Figure \ref{figsimulation}a displays the calculated curves of $R^0_{edge}$ and $R^0_{spine}$ in a 2D sample with aspect ratio matched to that in the experiment on Na$_3$Bi. As expected, with $\rho_{xx}$ set to a constant, the two curves diverge. This reflects the simultaneous enhancement of the $E$-field along the spine and its steep decrease at the edge caused by pure current jetting.

From the calculated resistances, we form the ratio 
\be
{\cal F}(u) = R^0_{edge}/R^0_{spine} = {\cal L}_e/{\cal L}_s. 
\label{eq:F}
\ee
The template curve ${\cal F}(u)$, which depends only on $u$, is plotted in semilog scale in Fig. \ref{figsimulation}b.

Turning to the values of $R_{edge}(B)$ and $R_{spine}(B)$ measured in Na$_3$Bi at the field $B$, we form the ratio ${\cal G}(B) = R_{edge}/R_{spine} = {\cal L}_e/{\cal L}_s$. Although ${\cal G}$ is implicitly a function of $u$, it is experimentally determined as a function of $B$ (how $u$ varies with $B$ is not yet known). We remark that ${\cal G}(B)$ and ${\cal F}(u)$ represent the same physical quantity expressed as functions of different variables. To find $u$, we equate ${\cal G}(B)$ to ${\cal F}(u)$ in the template curve. This process leads to the equation 
\be
u(B) = {\cal F}^{-1}({\cal G}(B)),
\label{eq:u}
\ee
from which we determine $u$ given $B$. Finally, because $R^0_{edge}$ and $R^0_{spine}$ are known from the simulation, we obtain the intrinsic profile of $\rho_{xx}$ as
a function of $B$ using the relations
\be
\rho_{xx}(B) = \frac{R_{edge}(B)}{R^0_{edge}(u)}\rho_0 = \frac{R_{spine}(B)}{R^0_{spine}(u)}\rho_0.
\label{rxx}
\ee
The redundancy (either resistance may be used) provides a useful check for errors in the analysis.

\begin{figure*}[t]
\includegraphics[width=16 cm]{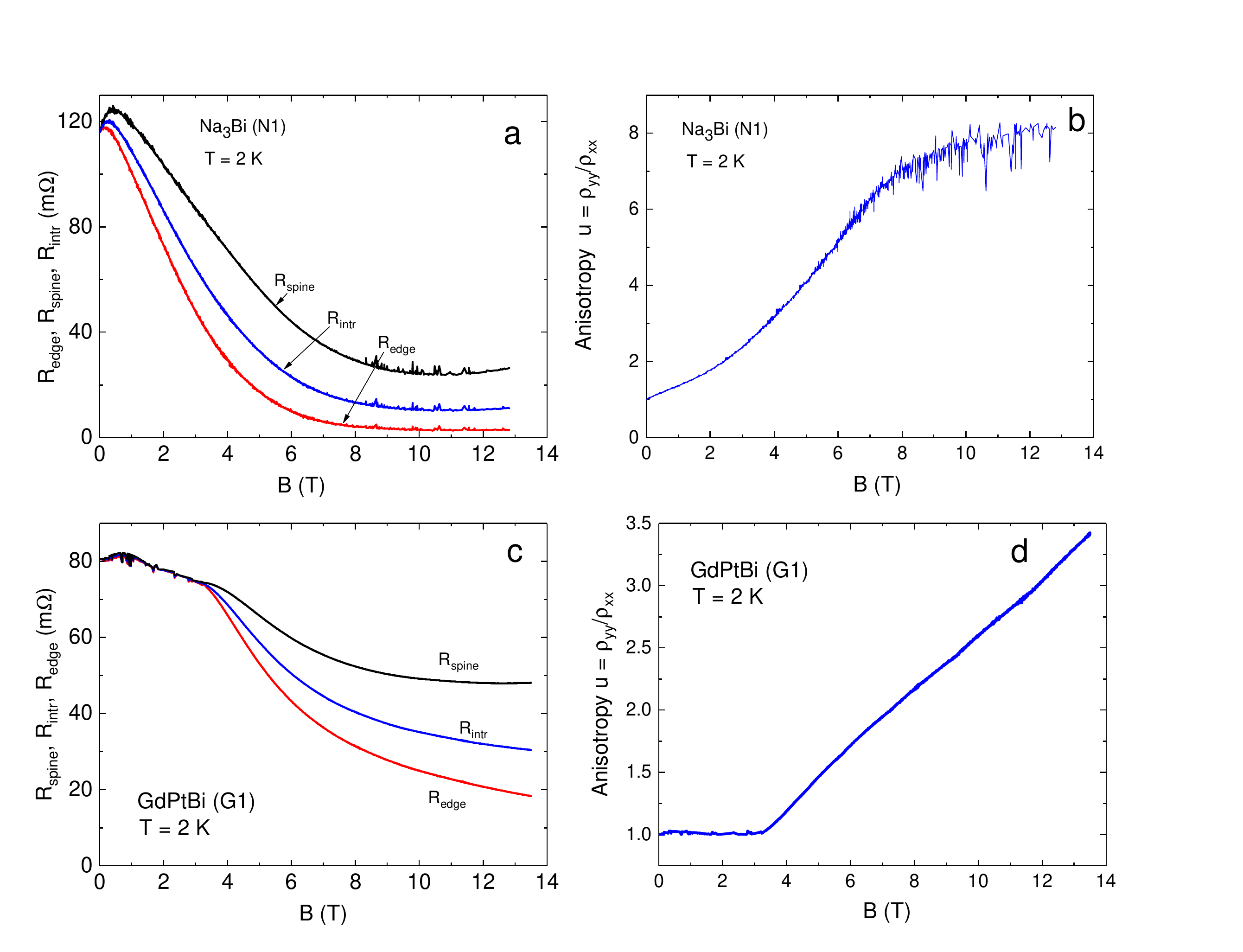}
\caption{\label{figintrinsic}
The intrinsic field profiles of $\rho_{xx}(B)$ and anisotropy $u(B)$ derived by applying Eqs. \ref{eq:F}, \ref{eq:u} and \ref{rxx}. Panel (a) shows the curves of $R_{spine}$ (black curve) and $R_{edge}$ (red) measured in Na$_3$Bi at 2 K. The inferred intrinsic profile $R_{intr}$ (blue curve) is sandwiched between the measured curves. In this sample (N1), $R_{intr}$ decreases by a factor of 10.9 between $B$ = 0 and 10 T. We note that $R_{spine}$ displays a weak minimum near 10 T which results from the competiting trends in $\rho_{xx}(B)$ and ${\cal L}_s(B)$. Panel (b) displays the intrinsic $u(B)$ in Na$_3$Bi derived from Eq. \ref{eq:u}. The corresponding curves for GdPtBi (Sample G1 at 2 K) are shown in Panels (c) and (d). Again, $R_{intr}(B)$ in Panel (c) is sandwiched between the curves of $R_{edge}(B)$ and $R_{spine}(B)$. Panel (d) plots the intrinsic anisotropy $u(B)$. In GdPtBi, the Weyl nodes only appear when $B$ is strong enough to force band crossing. In G1, this occurs at 3.4 T ($u$ remains close to 1 below this field).
}
\end{figure*}

The results of the analysis are shown in Fig. \ref{figintrinsic} for both Na$_3$Bi and GdPtBi.
In Fig. \ref{figintrinsic}a, we plot $\rho_{xx}(B)$ in Na$_3$Bi (as the intrinsic curve $R_{intr}(B)$, in blue). As expected, $R_{intr}(B)$ is sandwiched between the measured curves of $R_{edge}(B)$ and $R_{spine}(B)$. The field profile of the intrinsic anisotropy $u$ is displayed in Fig. \ref{figsimulation}d). 
It is interesting to note that, in Na$_3$Bi at 2 K, the intrinsic conductivity anisotropy increases to 8 as $B$ is increased to 14 T. This engenders significant distortion of $\bf J(r)$ away from uniform flow. The analysis provides a quantitative measure of how current jetting effects distort the measurements.  At 10 T, $R_{spine}$ is larger than $R_{intr}$ by a factor of 2.3 whereas $R_{edge}$ is 4.0$\times$ smaller than $R_{intr}$. The plots show explicitly how $R_{spine}(B)$ still decreases (by a factor of $\sim$5) between $B$ = 0 and 10 T, despite the enhancement in ${\cal L}_s$ caused by current jetting. Here we see explicitly that this occurs because the intrinsic LMR is so large (decreasing by by a factor of 10.9 between 0 and 10 T) that the current squeezing factor is always sub-dominant. With the procedure described, this sub-dominant distortion can be removed entirely. The corresponding profiles of $\rho_{xx}(B)$ and $u(B)$ in GdPtBi are shown in Fig. \ref{figintrinsic}c and \ref{figintrinsic}d, respectively. Unlike the case in Na$_3$Bi, a finite $B$ is required to create the Weyl nodes. This occurs at $\sim$3.4 T in G1. Below this field, the system is isotropic ($u$ close to 1). 

\section{Quantum vs. classical effects}\label{testscope}
From the experimental viewpoint, it is helpful to view the LMR experiment as a competition between the intrinsic anomaly-induced decrease in $\rho_{xx}$ (a quantum effect) and the distortions engendered by current jetting (classical effect). To observe a large, negative LMR induced by the chiral anomaly, it is imperative to have the chemical potential $\zeta$ enter the lowest Landau level LLL. The field at which this occurs, which we call $B_Q$, sets the onset field for this quantum effect. By contrast, the distortions to $\bf J$ caused by current jetting onsets at the field $B_{cyc}$, which is set by the inverse mobility $1/\mu$. We write $B_{cyc} = {\cal A}/\mu$ where the dimensionless parameter ${\cal A}= 5-10$, based on the numerical simulations (Eqs. \ref{eq:F} and \ref{eq:u}). 

\begin{figure*}[t]
\includegraphics[width=14 cm]{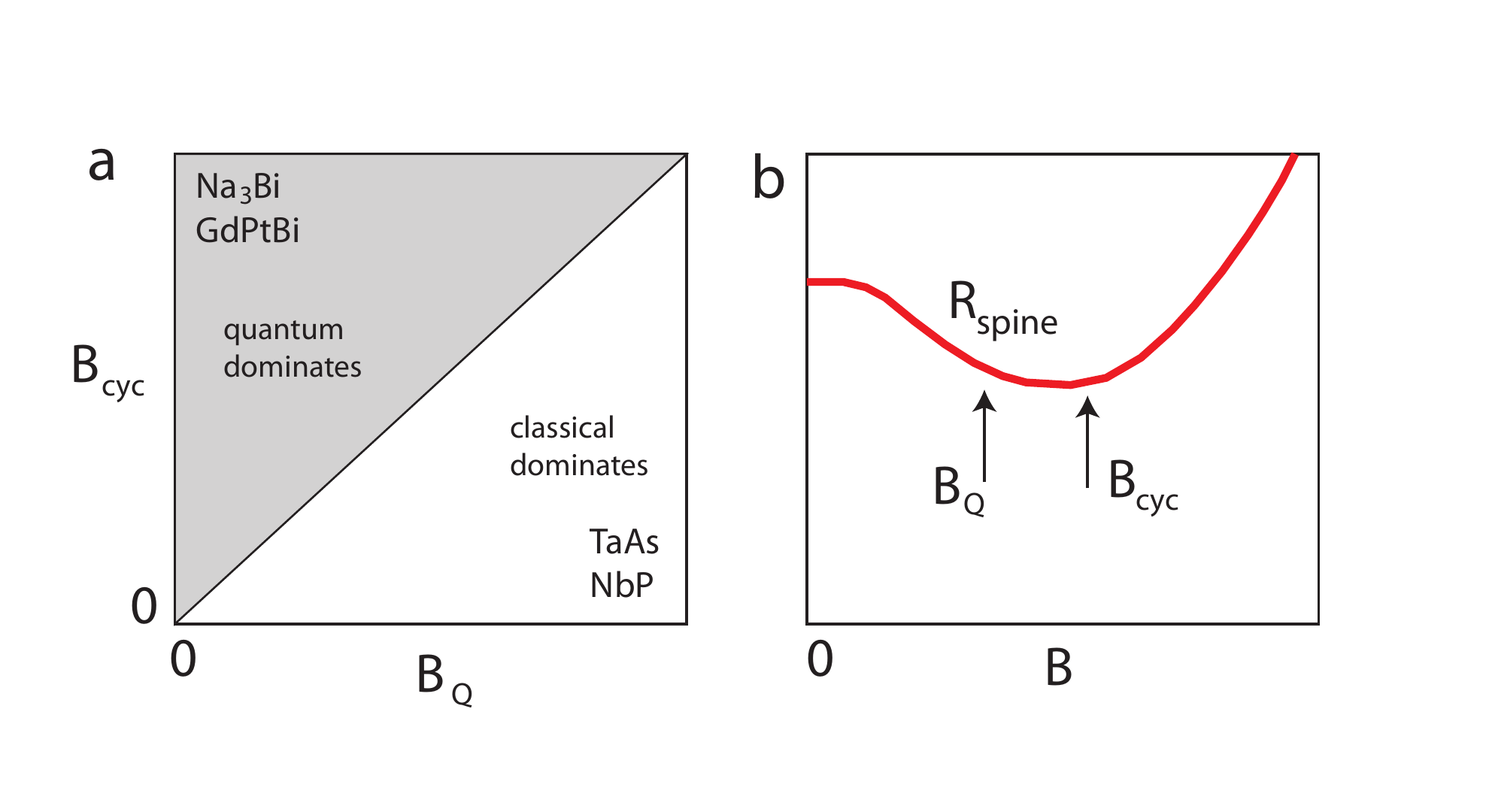}
\caption{\label{figscopetest}
Competition between the intrinsic chiral-anomaly LMR (with onset field $B_Q$) and the extrinsic classical effect of current jetting (onset field $B_{cyc} = {\cal A}/\mu$, ${\cal A}=5-10$). Panel (a): In the shaded upper half region, the system enters the LLL before the current jetting effects dominate as $B$ increases ($B_Q<B_{cyc}$). Panel (b) shows a hypothetical case close to the boundary $B_Q\simeq B_{cyc}$. In the profile of $R_{spine}$, an initial decrease is followed by a steep increase above $B_{cyc}$. For systems in the unshaded region ($B_Q>B_{cyc}$), the LLL is entered long after current jetting effects become dominant. The classical effect effectively screens the quantum behavior. Bulk crystals of Na$_3$Bi and GdPtBi fall safely in the upper half whereas TaAs and NbAs fall in the lower half. One way to avoid current jetting in the latter is to lower $B_Q$ by gating ultrathin-film samples.
}
\end{figure*}

If $B_Q<B_{cyc}$, the LLL is accessed before classical current distortion appears in increasing $B$. This is the situation in the upper shaded region in the $B_Q$ vs. $B_{cyc}$ space in Fig. \ref{figscopetest}a. The conditions are favorable for observing the chiral anomaly without worrying about classical current jetting. (To be sure, the chiral anomaly itself leads to a large anisotropy $\sigma_{xx}/\sigma_{yy}$ that can distort $\bf J$. However, this is a quantum effect that follows from the chiral anomaly and can be compensated for.) The measured curves of $R_{spine}$ and $R_{edge}$ bracket the intrinsic $\rho_{xx}$ which allows the latter to be obtained, as explained in Eqs. \ref{eq:F}, \ref{eq:u} and \ref{rxx}. In both Na$_3$Bi and GdPtBi, $B_Q\sim$ 5-6 T whereas $B_{cyc}$ exceeds 30 T. They fall safely within the shaded area. As $B_Q$ approaches $B_{cyc}$ (the diagonal boundary in Fig. \ref{figscopetest}a), classical current jetting becomes increasingly problematical. In Fig. \ref{figscopetest}b, the schematic curve illustrates the trend of how the quantum behavior can be swamped by the onset of current jetting. 

\begin{figure*}[t]
\includegraphics[width=12 cm]{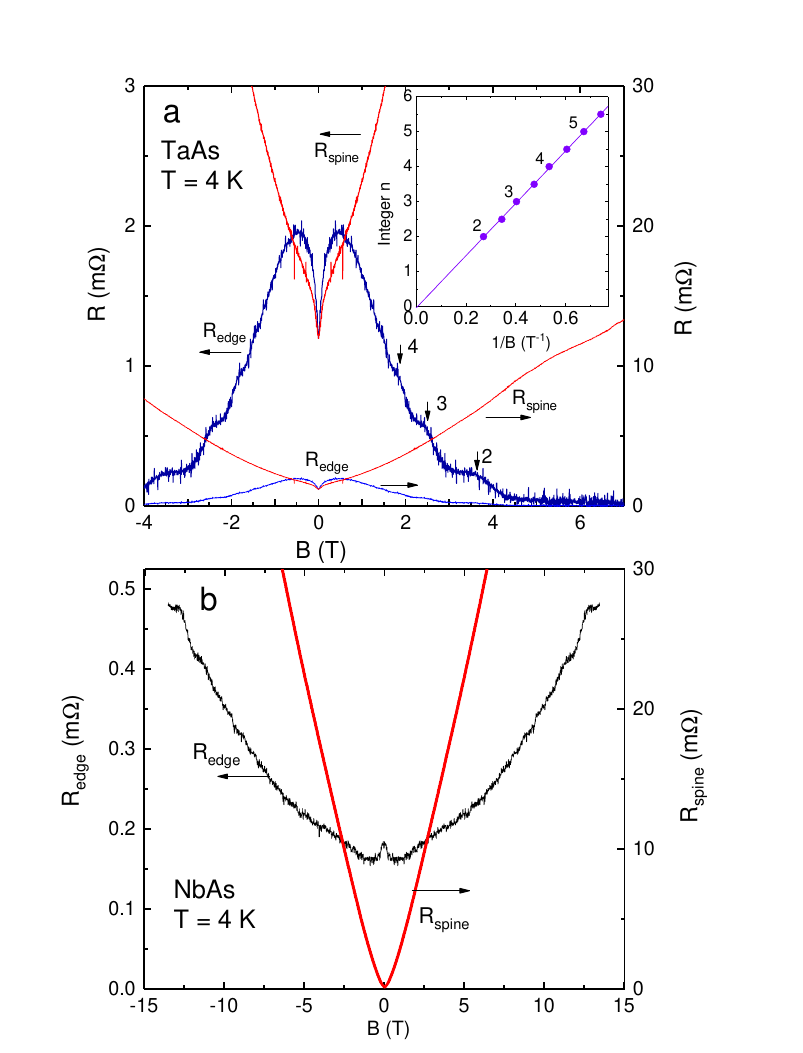}
\caption{\label{figWeyl}
The squeeze test applied to the Weyl semimetals TaAs and NbAs. Panel (a) shows field profiles of $R_{edge}$ and $R_{spine}$ in TaAs (measured at 4 K) in an in-plane, longitudinal $\bf B$ along the ``spine'' of the sample. The crystal is a thin plate of dimension $2.0\times 1.5\times 0.2$ mm$^3$ with spacing $\ell$ between voltage contacts of 1.5 mm, and current-contact diameters $d_c\sim$ 80 $\mu$m. Between 0.5 T and 6 T, $R_{edge}$ (thick blue curve, left axis) is observed to decrease by a factor of $\sim$50 to values below our resolution. However, the spine resistance $R_{spine}$ (red curve) increases by a factor of 7.8 (the right axis replots $R_{spine}$ and $R_{edge}$ reduced by a factor of 10). This implies that the pronounced decrease in $R_{edge}$ is an artifact caused by current jetting, just as observed in pure Bi. The Landau level indices $n$ of the weak SdH oscillations are indicated by vertical arrows. The index plot (inset) shows that $n$=1 is reached at $B_Q$= 7.04 T, whereas $1/\mu\sim$0.06 T. Panel (b) shows field profiles of $R_{edge}$ (black curve, left axis) and $R_{spine}$ (red curve, right axis) measured in NbAs at 4 K in longitudinal $\bf B$ (crystal size $1.5\times 1.0\times 0.4$ mm$^3$). The small negative LMR anomaly near zero $B$ in $R_{edge}$ has been invoked as evidence for the chiral anomaly even though $R_{edge}$ increases above 1.2 T. By contrast, $R_{spine}$ increases monotonically at a very steep rate (note difference in vertical scales). In both high-mobility semimetals, classical current-jetting effects onset near 0.4 T, well below $B_Q$. This makes the chiral anomaly virtually unobservable by LMR.
}
\end{figure*}

Finally, if $B_Q\gg B_{cyc}$, classical distortion effects onset long before the LLL is accessed. In this case, $R_{spine}$ and $R_{edge}$ display divergent trends vs. $B$. Even if an intrinsic LMR exists, we are unable to observe it in the face of the dominant (artifactual) change in $V_{xx}$ caused by classical current jetting. In the Weyl semimetals TaAs and NbAs, $B_Q\simeq$ 7 and 40 T, respectively, whereas $B_{cyc}\sim$0.4 T (because of their high mobilities). This makes LMR an unreliable tool for establishing the chiral anomaly in the Weyl semimetals. We illustrate the difficulties with $R_{edge}$ and $R_{spine}$ measured in TaAs (Fig. \ref{figWeyl}a) and in NbAs (Fig. \ref{figWeyl}b). 

In the initial reports, a weak LMR feature (5-10$\%$ overall decrease) was observed in TaAs and identified with the chiral anomaly. Subsequently, several groups found that both the magnitude and sign of the LMR feature are highly sensitive to voltage contact placement. We can in fact amplify the negative LMR to nearly 100$\%$. To apply the squeeze test, we have polished a crystal of TaAs to the form of a thin square plate and mounted contacts in the configuration sketched in Fig. \ref{figLandau}b, with small current contacts ($\sim 80\; \mu$m). As shown in Fig. \ref{figWeyl}a, $R_{edge}$ at 4 K (thick blue curve) displays a steep decrease, falling to a value approaching our limit of resolution at 7 T. Simultaneously, however, $R_{spine}$ (red) increases rapidly. The two profiles are categorically distinct from those in Na$_3$Bi and GdPtBi (Fig. \ref{figspine}c-f), but closely similar to the curves for Bi (Fig. \ref{figspine}a,b). Moreover, the weak SdH oscillations fix the field $B_Q$ needed to access the LLL at 7.04 T (inset in (a)). Since $1/\mu\sim$ 0.06 T, we infer that the classical current-jetting effect onsets long before the quantum limit is accessed. Hence TaAs is deep in the right-bottom corner of the phase diagram in Fig. \ref{figscopetest}a. The current jetting effects appear well before TaAs attains the quantum limit at $B_Q$, and completely precludes the chiral anomaly from being observed by LMR. 

Applying the squeeze test next to NbAs, we display the curves of $R_{edge}$ (black curve) and $R_{spine}$ (red) at 4 K in Fig. \ref{figWeyl}b. Here, both $R_{edge}$ and $R_{spine}$ increase with $B$, but $R_{spine}$ increases 100$\times$ faster (in the field interval 0$<B<$8.5 T, $R_{edge}$ doubles but $R_{spine}$ increases by a factor of 280). The vast difference in the rate of change is direct evidence for the squeezing of $J({\bf r})$ along the spine as depicted in Figs. \ref{figLandau}c. Again, with $B_Q\sim$40 T, we infer that NbAs falls deep in the right-bottom corner of Fig. \ref{figscopetest}a. Classical current jetting dominates the LMR. 

It is worth remarking that the squeeze test results do not invalidate the ARPES evidence showing that TaAs and NbAs are Weyl semimetals. Rather, they demonstrate that the negative LMR reported to date in the Weyl semimetals fall deep in the regime where classical current-jetting effects dominate.

Figure \ref{figscopetest}a suggests a way to avoid the screening effect of current jetting for the Weyl semimetals. By growing ultrathin films one may use gating to lower $\zeta$ towards zero in the Weyl nodes. This allows the LLL to be accessed at a much lower $B_Q$. Simultaneously, the increased surface scattering of the carriers will reduce $\mu$ (hence increase $B_{cyc}$). In allowing the quantum effect to onset before the classical effect becomes dominant, both trends shift the ``operating point'' towards the shaded region $B_Q<B_{cyc}$. The ability to tune $B_Q$ by gating will enable more tests for mapping out the current density distribution. The squeeze test is actually easier to implement using thin-film samples. Because several groups worldwide are attempting to grow thin-film TaAs and NbAs, the prospects for the Weyl semimetals seem quite encouraging.

A fourth candidate for the chiral anomaly is ZrTe$_5$~\cite{QLi,Liang}, which displays a moderately large negative LMR signal ($\rho_{xx}$ decreases by 35$\%$). A very recent experiment~\cite{Liang} has detected a (true) planar Hall effect when the chiral anomaly appears. This implies the simultaneous appearance of a large Berry curvature in applied $\bf B$. However, currently available bulk crystals have a narrow ribbon-like morphology unsuited for the squeeze test (a platelike shape is optimal). However, using focussed ion beam (FIB) techniques, we may envisage sculpting the ribbons into thin plates. Microlithography techniques can then be harnessed to deposit voltage contacts for measuring $R_{spine}$ and $R_{edge}$. We are not aware of any technical barrier that would preclude applying the squeeze test on plate-like crystals tens of $\mu$m on a side. The field profiles of $R_{spine}$ and $R_{edge}$ may then be compared as reported here in both Na$_3$Bi and GdPtBi. The FIB technique can be applied to future chiral-anomaly candidate materials that do not readily grow as large crystals.

\section{Planar Angular Magnetoresistance}\label{AMR}
As shown in Eqs. \ref{R1} and \ref{R2}, the growth of the anisotropy $u$ arises differently in Cases (1) and (2). The difference leaves a strong imprint on the planar angular magnetoresistance (AMR) which we describe here. In an AMR experiment, $\bf B$ is rotated within the $x$-$y$ plane while the longitudinal and transverse voltages are recorded. AMR experiments have been used to investigate the resistivity anisotropy produced by the magnetization $\bf M$ in ferromagnetic thin films. Recently, Burkov~\cite{Burkov17} has suggested that AMR measurements may be used to probe the chiral anomaly.

The sample geometry is as defined above, but now with broad current contacts and a pair of standard Hall contacts spaced along $y$ (see inset in Fig. \ref{figplanar}b). The lab frame ($x$ and $y$ axes) remain fixed to the sides of the sample. The in-plane $\bf B$ determines the sample's orthogonal frame $\bf a$, $\bf b$ and $\bf c$ ($\bf a\parallel B$ is tilted at an angle $\theta$ relative to $\bf\hat{x}$ with $\bf c\parallel \hat{z}$). The tilt produces potential drops $V_{xx}$ and $V_{yx}$ given by
\begin{eqnarray}
V_{xx}/I &=& \rho_{bb} + \Delta\rho \cos^2\theta, \label{Vxx}\\ 
V_{yx}/I &=& \Delta\rho \sin\theta\cos\theta, \label{Vyx}
\end{eqnarray}
where $\rho_{aa}$ and $\rho_{bb}$ are the resistivities measured along axes $a$ and $b$, respectively, and $\Delta\rho = \rho_{aa} - \rho_{bb}$.

[By convention, the transverse voltage $V_{yx}$ is dubbed the ``planar Hall effect'' even though it is strictly even in $\bf B$. As $V_{yx}$ does not satisfy the Onsager relation for a true Hall response, this is a misnomer. (In topological matter, the Berry curvature can generate a true in-plane Hall signal that is odd in $\bf B$ and distinct from $V_{yx}$ in Eq. \ref{Vyx}.) To avoid confusion, we call $V_{yx}$ the off-diagonal AMR signal, and $V_{xx}$ the longitudinal AMR signal.]

Generally, the AMR results are not very informative (the same angular pattern is obtained regardless of the microscopic origin of the anisotropy). However, for our problem, we find that the parametric plot of $V_{yx}$ vs. $V_{xx}$ provides a litmus test that distinguishes Case 1 from Case 2.

In Case (1), with $\theta=0$ ($\bf B\parallel\hat{x}$), $V_{xx}$ detects $\rho_{aa}{\cal L}_e$; its ``spurious'' decrease as $B$ increases arises entirely from ${\cal L}_s$. In the orthogonal situation $\theta = \pi/2$, $V_{xx}$ detects $\rho_{bb}{\cal L}_0$ (i.e., beam focussing effects are absent). By juxtaposition, the two measurements reveal how $u$ behaves (see Eqs. \ref{R1}). This is best shown by plotting $V_{yx}$ against $V_{xx}$ with $\theta$ as the running parameter at a fixed value of the magnitude $B$. In weak $B$, the contours describe small loops circling the zero-$B$ point. As $B$ increases, they expand dramatically away from the zero-$B$ point in the direction of increasing $V_{xx}$. This lop-sided expansion (resembling a shock-wave) reflects the sharp increase in the resistivity $\rho_{bb}$ measured orthogonal to $\bf B$ (while $\rho_{aa}$ remains unchanged; see Eq. \ref{rhoab}). Indeed, from Eq. \ref{rhoab}, we have in the high-$B$ limit
\be
\rho_{bb} \to \frac{(\mu_e\mu_h)^2B^2}{(\sigma_h\mu_e^2 + \sigma_e\mu_h^2)}.
\label{rhob}
\ee
$\rho_{bb}$ increases as $B^2$ without saturation. Hence, in Case 1, we expect the caliper of the contours (given by $\Delta\rho$) to expand without limit as $B^2$.

Case (2) yields a qualitatively different parametric plot. In the chiral anomaly regime, our measurements show that $\rho_{aa}$ (captured by $V_{xx}$ at $\theta=0$) decreases intrinsically with increasing $B$, while $\rho_{bb}$ (at $\theta=\pi/2$) increases by roughly the same fraction. The balanced changes lead to closed contours that expand roughly isotropically from the zero-$B$ point. Moreover, the contour calipers $\Delta\rho$ approach saturation at large $B$. 

\section{Parametric plots}\label{param}
\begin{figure*}[t]
\includegraphics[width=16 cm]{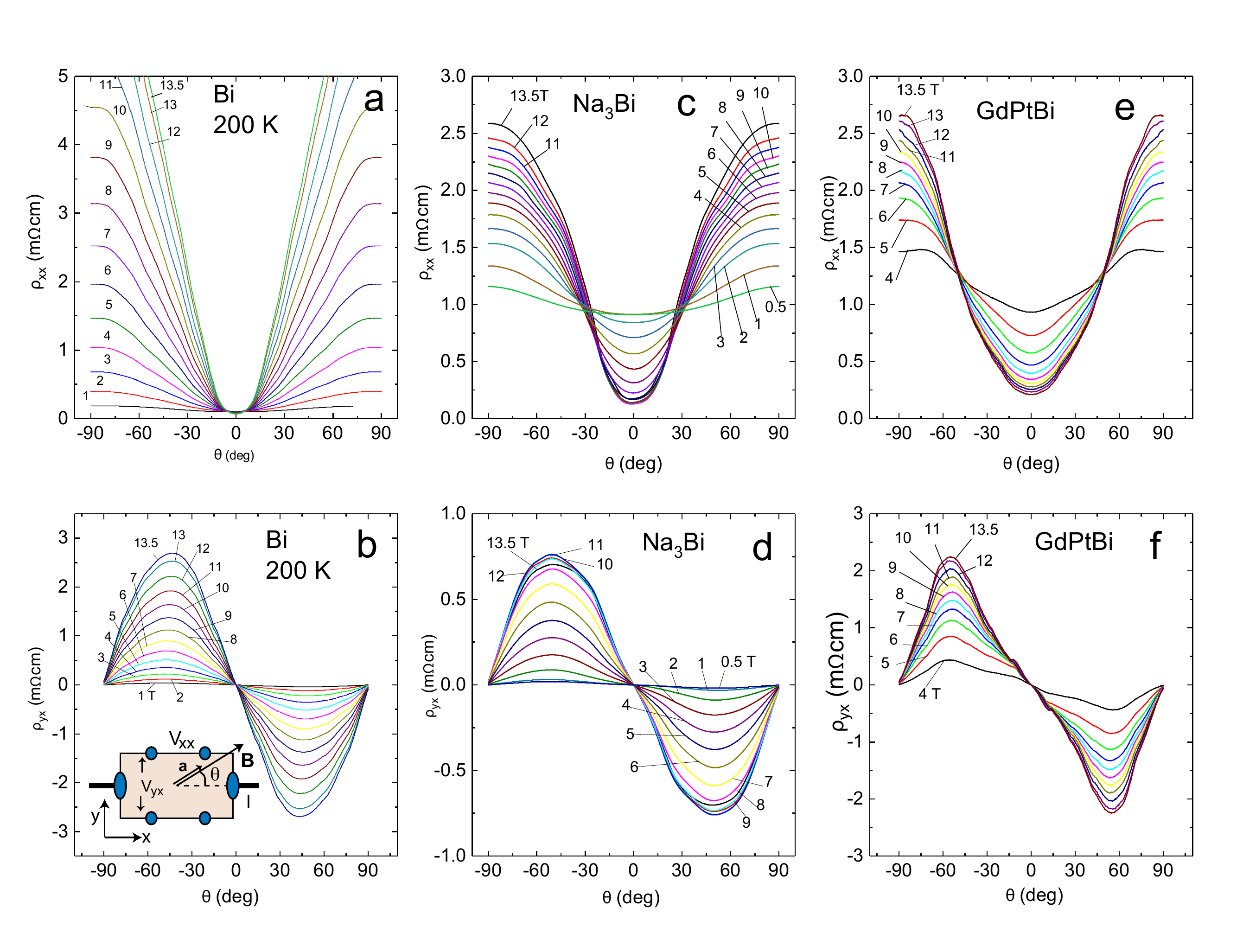}
\caption{\label{figplanar} 
Planar angular magnetoresistance (AMR) measurements in Bi, Na$_3$Bi and GdPtBi. Panels (a) and (b) show the $\theta$ dependence of the diagonal and off-diagonal AMR signals, expressed as effective resistivities $\rho_{xx}$ and $\rho_{yx}$, respectively, in Bi (Sample B2) at 200 K. In Panel (a), $\rho_{xx}$ decreases with increasing $B$ when $\theta = 0$ (spurious LMR produced by current jetting). When $|\theta|$ exceeds 20$^\circ$, $\rho_{xx}$ increases steeply with both $|\theta|$ and $B$. In absolute terms, the increase in $\rho_{xx}$ at $\theta = 90^\circ$ far exceeds its decrease at $\theta = 0$. The off-diagonal MR $\rho_{yx}$ follows the standard $\sin\theta.\cos\theta$ variation (Panel (b)).  Panel (c) shows curves of $\rho_{xx}$ for Na$_3$Bi (Sample N2) at 4 K. The decrease in $\rho_{xx}$ at $\theta = 0$ is roughly comparable with the increase at $90^\circ$. A similar balanced pattern is observed in GdPtBi (Sample G1) measured at 2 K (Panel (e)). In both Case (2) systems, the comparable changes in $\rho_{xx}$ measured at 0$^\circ$ and 90$^\circ$ contrast with the lop-sided changes seen in Bi (Case (1)). The off-diagonal $\rho_{yx}$ in both Na$_3$Bi and GdPtBi (Panels (d) and (f)) show the standard $\sin\theta.\cos\theta$ profile. The profiles in Panel (f) are distorted by anisotropic features associated with the creation of the Weyl pockets. The inset in Panel (b) shows the sample geometry with $\bf a\parallel B$, both at the tilt angle $\theta$ to $\bf\hat{x}$.
}
\end{figure*}

As in Sec. \ref{squeeze results}, we compare the planar angular MR results in the 3 materials, pure Bi, Na$_3$Bi and GdPtBi. Figure \ref{figplanar}a displays the angular profiles $\rho_{xx}$ vs. $\theta$ at selected field magnitudes $B$ measured in Bi at 200 K with $\bf J\parallel \hat{x}$. As $\bf B$ is tilted away from alignment with $\bf J$ ($\theta = 0$), $\rho_{xx}$ increases very rapidly at a rate that varies nominally as $B^2$. The overall behavior in $\rho_{xx}$ is a very large increase with $B$ as soon as $|\theta|$ exceeds 10$^\circ$. However, at $\theta = 0$, a decrease in $\rho_{xx}$ of roughly 50$\%$ can be resolved. This is the spurious LMR induced by pure current jetting. The off-diagonal signal $\rho_{yx}$ shows the $\sin\theta.\cos\theta$ variation described in Eq. \ref{Vyx} ($\rho_{yx}$ is strictly even in $B$). 

The corresponding traces of $\rho_{xx}$ and $\rho_{yx}$ measured in Na$_3$Bi at 2 K are shown in Figs. \ref{figplanar}c and \ref{figplanar}d, respectively. Although the curves for $\rho_{yx}$ are similar to those in Bi, a qualitatively different behavior in $\rho_{xx}$ becomes apparent. At $\theta=0$, $\rho_{xx}$ is suppressed by a factor of $\sim 7$ (when the chiral anomaly appears). In the transverse direction ($\theta = 90^\circ$), the poor conductance transverse to $\bf B$ in the LLL raises $\rho_{xx}$ by a factor $\sim 2.5$. In terms of absolute magnitudes, the changes to $\rho_{xx}$ are comparable along the two orthogonal directions, in sharp contrast with the case in Bi. This ``balanced'' growth leaves a clear imprint in the parametric plots. The off-diagonal signal $\rho_{yx}$ displays the same $\sin\theta.\cos\theta$ variation as in Bi. 

The plots of $\rho_{xx}$ and $\rho_{yx}$ for GdPtBi in Figs. \ref{figplanar}e and \ref{figplanar}f also show a concentric pattern. A complication in GdPtBi is that the nature of the Weyl node creation in field (by Zeeman shift of parabolic touching bands) is anisotropic (dependent on the direction of $\bf B$). The existence of low-$B$ oscillations adds a modulation to the off-diagonal curves, which distorts the variation from the $\sin\theta.\cos\theta$ form. Nonetheless, a balanced growth in $\rho_{xx}$ is also observed (Fig. \ref{figplanar}e).

\begin{figure*}[t]
\includegraphics[width=16 cm]{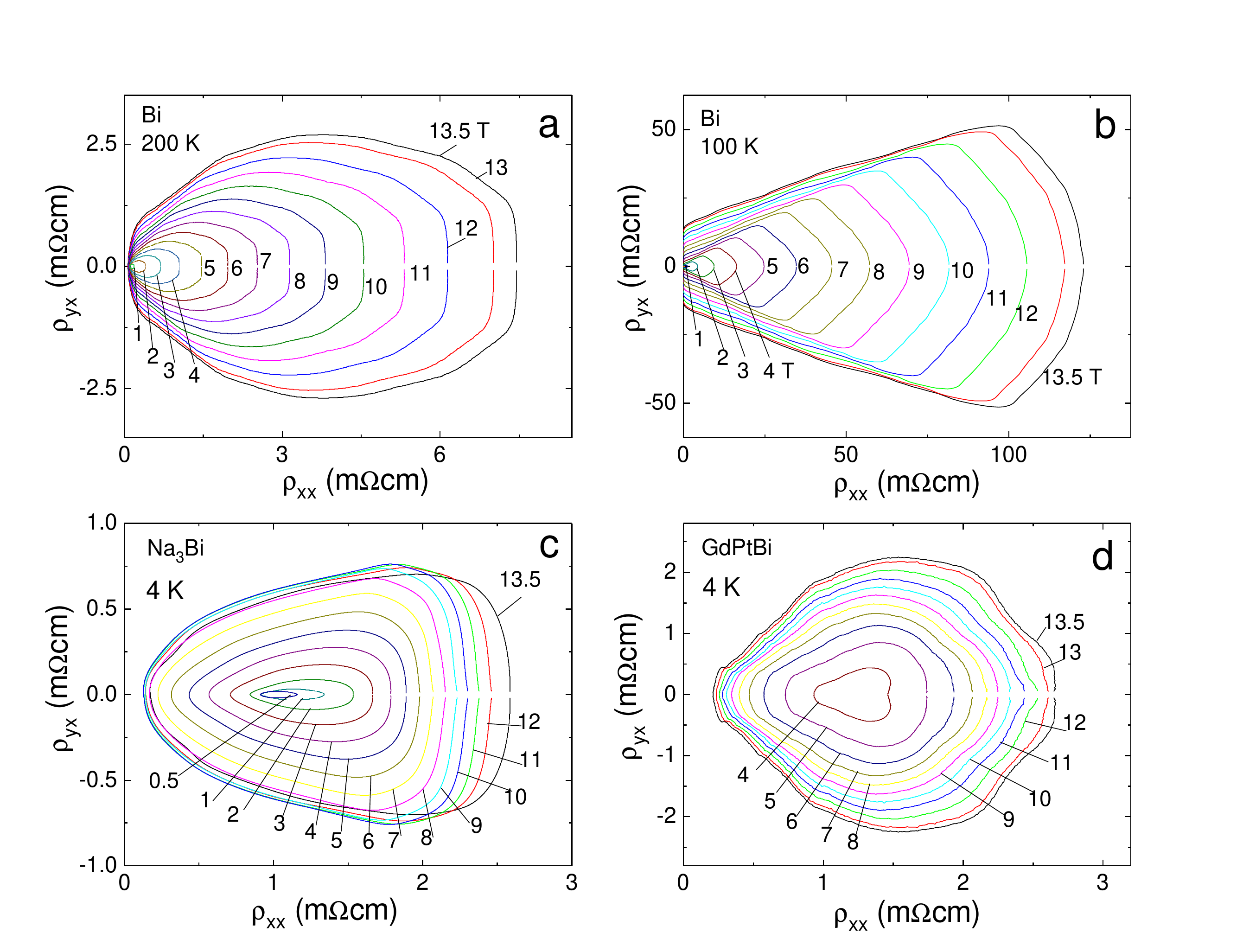}
\caption{\label{figparam} 
Parametric plots of the planar AMR signals. Panel (a) shows the orbits in pure Bi at 200 K (Sample B2) obtained by plotting $\rho_{xx}$ vs. $\rho_{yx}$ with $\theta$ as the parameter and $B$ kept fixed. The orbits start at the left (small $\rho_{xx}$) and end at the right (large $\rho_{xx}$). As $B$ increases, the orbits expand without saturation to the right. Panel (b) shows the same plots measured at 100 K. The increased mobility strongly amplifies the lop-sided expansion pattern, which resembles a shock-wave. The orbits in the parametric plots measured in Na$_3$Bi (Sample N2, at 4 K) and GdPtBi (G1) at 2 K [Panels (c) and (d), respectively] are nominally concentric around the point at $B$ = 0. The differences between Case (1) and Case (2) reflect the very different behaviors of $\rho_{xx}$ at the extreme angles $\theta = 0$ and $90^\circ$, as discussed in Fig. \ref{figplanar}.
}
\end{figure*}

Figure \ref{figparam} compares the parametric plots in Bi at $T$ = 200 K (Panel a) and 100 K (b). In each orbit ($B$ set at indicated value), $\theta$ starts at 0$^\circ$ on the left limb and ends at $90^\circ$ on the right. In Fig. \ref{figparam}a, the steep increase on the right limb causes the orbits to expand strongly to the right. At 100 K (Panel b), the higher mobility creates exaggerated skewing of the rightward expansion leading to the emergence of a ``shock-wave'' pattern. By contrast, the parametric plots in both Na$_3$Bi (Fig. \ref{figparam}c and GdPtBi (Panel d) show concentric orbits that expand in a balanced pattern as anticipated above. The contrast between Case (1) [Panels (a) and (b)] and Case (2) reflects directly the distinct nature of mechanisms that increase the anisotropy $u$ (see Eqs. \ref{R1} and \ref{R2}). 

\section{Concluding remarks}\label{conclude}
As mentioned in Sec. \ref{intro}, the existence of a negative LMR (longitudinal magnetoresistance) that is intrinsic is relatively rare. However, the observation of an artifactual decrease $V_{xx}$ in the LMR geometry is a common experience in high mobility semimetals. To assist in the task of disentangling the rare intrinsic cases from extrinsic cases (mostly caused by current jetting), we described a test that determines the current-jetting distortions and a procedure for removing them. The squeeze test consists of comparing the effective resistance $R_{spine}$ measured along the spine (line joining current contacts) with that along an edge $R_{edge}$. In pure Bi, $R_{spine}$ increases dramatically with $B$, while $R_{edge}$ decreases. However, in both Na$_3$Bi and GdPtBi, both $R_{spine}$ and $R_{edge}$ decrease. Hence an inspection of the trend in $R_{spine}$ allows Case (1) (here Bi) to be distinguished from Case (2), the two chiral-anomaly semimetals. From the factorization implicit in Eqs. \ref{Re} and \ref{Rs}, we relate the experimental ratio ${\cal G}$ to the curve ${\cal F}$ obtained by numerical simulation. This enables the intrinsic profiles of both $\rho_{xx}(B)$ and $u$ to be obtained from measurements of $R_{edge}$ and $R_{spine}$. After removal of the distortions, $\rho_{xx}$ is seen to decrease by a large factor (10.9) between $B$ = 0 and 10 T, while $u$ increases by ~8. (For simplicity, the numerical simulation was done in the 2D limit which we judge is adequate for flake-like samples. Obviously, this can be improved by adopting a fully 3D simulation.) The yes/no nature of the test based on inspection of $R_{spine}$, bolstered by the quantitative analysis which removes the sub-dominant corrections, adds considerable confidence that the chiral-anomaly LMR profiles in Na$_3$Bi and GdPtBi are intrinsic. Moreover, the subdominant distortion factors arising from current jetting can be effectively removed. 

In Sec. \ref{testscope}, we described LMR experiments as a competition between the intrinsic quantum effect arising from the chiral anomaly and the classical effects of current jetting. The former describes a phenomenon intrinsic to massless chiral fermions. To see it in full force, the applied $B$ should exceed $B_Q$, the field needed to move $\zeta$ into the LLL. This point seems worth emphasizing because in many reports the claimed anomaly seems to appear in weak fields $B\ll B_Q$. The experimental concern is that once current jetting appears (at the field scale $B_{cyc}$) it inevitably engenders a dominant, negative LMR profile that is extrinsic in origin. The divergent field profiles of $R_{spine}$ and $R_{edge}$ provide a strong warning that the LMR profile is then highly unlikely to be intrinsic. 

Looking ahead, we discussed in Sec. \ref{testscope} how the classical screening effect from current jetting may be avoided by using ultrathin, gateable films of the Weyl semimetals which may become available in the near future. For the fourth class of chiral anomaly semimetal ZrTe$_5$~\cite{QLi,Liang}, we propose using a focussed ion beam to sculpt platelike, samples 10 $\mu$m on a side, and applying microlithography to attach contacts for the squeeze test.


\newpage

\vspace{1cm}\noindent

$^\dagger$Corresponding author's email: npo@princeton.edu

\vspace{5mm}\noindent
{\bf Acknowledgements} 
The research was supported by the U.S. Army Research Office (W911NF-16-1-0116) and a MURI award for topological insulators (ARO W911NF-12-1-0461). NPO acknowledges support from the Gordon and Betty Moore Foundation's Emergent Phenomena in Quantum Systems Initiative through Grant GBMF4539. The growth and characterization of crystals were performed by S.K. and R.J.C., with support from the National Science Foundation (NSF MRSEC grant DMR 1420541). Work at UCLA was supported by the U.S. Department of Energy (DOE), Office of Basic Energy Sciences, Award DE-SC0011978.

\vspace{3mm}
\noindent

\newpage

\end{document}